\documentclass[aps,prstab,preprint,groupedaddress]{revtex4}
\usepackage{graphics,epsfig}
\begin{document}

\draft
\title{Enhanced Winning in a Competing Population by Random Participation}
\author{K.F. Yip$^{1}$, T.S. Lo$^{1}$, P.M. Hui$^{1}$, and N.F. Johnson$^{2}$}
\address{$^{1}$ Department of Physics, The Chinese University of
Hong Kong\\ Shatin, New Territories, Hong Kong}
\address{$^{2}$ Physics Department, University of Oxford\\ Clarendon
Laboratory, Oxford OX1 3PU, United Kingdom}


\begin{abstract} We study a version of the
minority game in which one agent is allowed to join the game in a
random fashion.  It is shown that in the crowded regime, i.e., for
small values of the memory size $m$ of the agents in the
population, the agent performs significantly well if she decides
to participate the game randomly with a probability $q$ {\em and}
she records the performance of her strategies only in the turns
that she participates.  The information, characterized by a
quantity called the inefficiency, embedded in the agent's
strategies performance turns out to be very different from that of
the other agents. Detailed numerical studies reveal a relationship
between the success rate of the agent and the inefficiency. The
relationship can be understood analytically in terms of the
dynamics in which the various possible histories are being visited
as the game proceeds. For a finite fraction of randomly
participating agents up to 60\% of the population, it is found
that the winning edge of these agents persists.

\vspace*{0.1 true in}


\end{abstract}

\maketitle
\newpage

\section{Introduction}

The self-organization of an evolving population consisting of
agents competing for limited resource is an important problem in
the science of complex systems and has potential applications in
areas such as economics, biological, engineering, and social
sciences\cite{recentactivities,ourbook}. The bar-attendance
problem proposed by Arthur \cite{arthur,johnson1}, for example,
constitutes a typical setting of such system in which a population
of agents decide whether to go to a bar with limited seating
capacity for pleasure.  The agents are informed of the attendance
in past weeks, and hence the agents share common information,
interact through their actions, and learn from past experience.
The problem can be simplified by considering binary games, either
in the form of the minority game (MG) \cite{challet1} or in a
binary-agent-resource (B-A-R) game \cite{johnson2}. For low
resource level in which there are more losers than winners, the
minority Game proposed by Challet and Zhang \cite{challet1}
represents a simple, yet non-trival, model of competing
populations.

The MG comprises of an odd number $N$ of agents. At each time
step, the agents independently decide between option ''$0$'' and
option ''$1$''$.$ The winners are those who have chosen the
minority side.  The agents learn and adapt by evaluating the
performance of their strategies, which map the global information,
i.e., records of the most recent $m$ winning options, to an
action. A characteristic quantity in MG is the standard deviation
$\sigma$ of the number of agents making a particular choice. This
quantity reflects the performance of the population as a whole in
that a small $\sigma $ implies on average more winners per turn,
and hence a higher success rate per turn per agent. In MG,
$\sigma$ exhibits a non-monotonic dependence on the memory size
$m$ of the agents \cite{savit,challet3,challet4}. When $m$ is
small, there is much overlap between the agents' strategies. This
crowd effect \cite{johnson2,johnson3,hart1} leads to a large
$\sigma$, implying the number of losers is high. This is the
crowded or efficient phase of MG. In the inefficent phase where
$m$ is large, $\sigma$ is moderately small and the agents perform
better than deciding randomly.

Two important questions in MG and other multi-agent based models
are that (i) how one can possibly suppress $\sigma$ and lead to a
better performance of population as a whole, and (ii) how an
individual agent can possibly out-perform other agents without
giving the agent too much extra capability.  The former is clearly
crucial from the point of view of a governing body in improving a
society's performance, and the latter is important from the
viewpoint of the agents.  Much attention on the MG have been
focused on the first question, and several models with a
suppressed $\sigma$ have been proposed. The thermal model of MG
\cite{cavagna}, for example, allows the agents to use strategies
other than the best scoring one with a probability taking on the
form of a Boltzmann factor. It was found that $\sigma$ can be
suppressed significantly in the crowded phase, as the
probabilistic usage of strategies reduces the crowd effect
\cite{hart2}. Alternatively, the use of a personal information
\cite{person} for decisions, e.g., based on each agent's record of
her own actions, instead of a globally announced information
provides another mechanism for suppressing the crowd effect.  In
the evolutionary minority game proposed by the present
authors\cite{emg1,emg2}, the agents can adapt their behavior by
adjusting their probability of following a strategy common to all
agents. It was shown that the agents would self-organize in such a
way that $\sigma$ is small. Recently, Sysi-Aho {\em et al.}
\cite{gene1} proposed an genetic algorithm for the evolution of
the agents and found that the fluctuations are reduced. A common
feature of these models is that the way through which the agents
adapt to past performance is significantly altered, as compared
with the MG.

In the present work, we focus on the question of how individual
agents may out-perform their competitors. Our model is motivated
by realistic behavior of ordinary people.  Taking the situation of
an agent buying or selling stocks, for example. He may not enter
the market to sell or buy stocks every day. In addition, people
tend to learn from their hands-on experience, i.e., they tend to
learn in the turns that they actually participate. When they
trade, they decide based on updated information.  Based on these
common observations, we propose a model in which an agent or a
fraction of agents participate in the MG with a probability $q$
per turn and these agents register the performance of their
strategies {\em only} in the turns that they participate.
Strikingly, it is found that the winning probability of these
randomly participating agents (henceforth referred to as RPA) are
significantly enhanced, compared with the other agents who enter
the MG every turn.

The plan of the paper is as follows. In section II, we introduce
our model. Results of extensive numerical studies for one randomly
participating agent with $q=0.5$ in a population are presented in
Sec.III.  To explore the underlying physics of the enhanced
success rate of the agent, we study the statistics of the
bit-strings observed by the agent.  Results on the information
observed, which is described by a quantity called the inefficiency
$\varepsilon $, are presented in Sec.IV. It is found that a
non-trivial relationship exists between the success rate of the
agent in a run with $\varepsilon$ of that run. A theoretical
analysis on the observed relationship is also given in Sec.IV. In
Sec.V, we present numerical results for $q\neq 0.5$ and generalize
our discussion on the relationship between the success rate and
$\varepsilon $ to arbitrary values of $q$. Results for many RPAs
in a population are presented in Sec.VI, together with results
comparing the performance of RPAs and random agents who
participate and decide randomly.  It is found that the RPAs, with
their scheme of assessing their strategies, consistently perform
better than random agents. Section VII gives a summary of the
results.

\section{The Model}

The basic MG \cite{challet1} comprises of $N$ agents competing to
be in a minority group at each time step. The only information
available to the agents is the history. The history is a
bit-string of length $m$ recording the minority option for the
most recent $m$ time steps. There are a total of $2^{m}$ possible
history bit-strings. At the beginning of the game, each agent
picks $s$ strategies, with repetition allowed. They make their
decisions based on their strategies. A strategy is a look up table
with $2^{m}$ entries giving the predictions for all possible
history bit-strings. Since each entry can either be `0' or `1',
the whole strategy pool contains $2^{2^{m}}$ strategies.
Adaptation is built in by allowing the agents to accumulate a
merit (virtual) point for each of her strategies as the game
proceeds, with initial merit point set to zero for all strategies.
After each turn, one (virtual) point is assigned to the strategies
that would have predicted the winning minority option. At each
turn, the agent follows the prediction of her best-scoring
strategy.  A random choice will be made for tied strategies.

In the present model, we consider a population of $N$ agents in
which there are a number $N_{RPA}$ of randomly participating
agents. These agents are allowed to join the game with a
probability $q$, i.e., a RPA has a probability $q$ of joining the
game in each turn and a probability of $1-q$ of staying out of the
game in a turn. The other ($N-N_{RPA}$) agents, as in the MG,
participate every turn. For all agents, they decide in the same
way as in the MG when they participate, i.e., they decide based on
the best-scoring strategy that they hold at the moment of decision
and follow the prediction of the strategy for the given most
recent $m$ winning options. In the case that the two possible
options are chosen by the same number of agents, a case that may
occur when RPAs are present, the winning outcome is decided
randomly.

The RPAs differ from the other agents in the following way.
Besides joining the game with probability $q$, they reward the
virtual points to the strategies that they hold {\em only} in the
turns that they participate.  For the turns that a RPA decides not
to play, no virtual points are awarded to her strategies,
regardless of the outcome.
These RPAs, therefore, carry the typical features of ordinary
people as described in the previous section. For $q=1$ or
$N_{RPA}=0$, the present model reduces to the MG.

\section{Results for one Randomly Participating Agent:
$\lowercase{q}=0.5$}

We have performed extensive numerical simulations on our model.
First, we study the case of one RPA in a population with $q=0.5$
in which the RPA is participating randomly, e.g. deciding by
tossing a unbiased coin.  We consider systems of $N=101$ and $301$
agents, with each agent holding $s=2,3$ strategies. The quantity
of interest is the success rate, which is the winning probability
of the agents. For the RPA, the success rate $R$ is the ratio of
the number of winning turns to the number of turns she has
actually participated.

Figure 1 shows the success rate of the RPA, together with the
success rate of the other ($N-1$) agents as a function of the
memory size $m$ for a system with $N=101$ agents for $s=2$
(Fig.1(a)) and $s=3$ (Fig.1(b)). For each value of $m$, 50
independent runs with different initial random distributions of
strategies among the agents are carried out. The lines represent
an average over the 50 runs. The most striking feature of the
results is that in the crowded phase where $m$ is small, i.e., $2
\cdot 2^{m} < N\cdot s$, the RPA performs {\em significantly
better} than the agents who participate every turn. The spread of
results in different runs is small and hence the enhanced success
rate of the RPA is an intrinsic feature. For higher values of $m$
in the inefficient phase of the MG, the success rate of the RPA is
found to vary much from run to run, with an average success rate
lower than that of the other agents. The crossover from one
behavior to another occurs at a value $m_{o}$, which takes on a
value close to the crossover from the crowded phase to the
inefficient phase in the basic MG. For very high values of $m$,
the success rates of the RPA and the rest of the population become
identical.  Similar behavior is also found for $s=3$.  Comparing
Fig.1(a) and Fig.1(b), the enhanced success rate of the RPA over
the other agents in the range of small $m$ is more pronounced for
higher values of $s$, as the population does not perform
collectively well in the basic MG as $s$ increases for small $m$.

To test whether the dependence of success rate of the RPA on $m$
is intrinsic, we carried out numerical simulation for a system
with $N=301$ with $s=2,3$. The success rate of a  RPA, averaged
over 50 runs, is shown in Fig. 2, together with the results for
$N=101$ and $s=2$ for comparison. In general, the average success
rate depends only weakly on $m$ for $m<m_o$. Note that the value
of $m_o$ depends on both $N$ and $s$. Around $m_o$, the RPA's
success rate drops and reaches a minimum before increasing with
$m$ again for $m>m_o$.

Qualitatively, the enhanced winning of a RPA comes from a
successful escape in over-adapting to the history created by
action of the other agents. In the MG, adaptation is achieved by
assessing the performance of each strategy as the game proceeds.
In the crowded phase, it has been shown that the history
bit-string exhibits features with a periodicity of length
$2\times2^{m}$ \cite{zheng}. For small values of $m$, agents in
the population adapt too effectively to the history. When a
particular history bit string occurs for the first time (or has
occurred for an even number of times in previous turns), the
agents basically decide randomly. The outcome then leads to
virtual points being awarded to those strategies predicted the
outcome for that particular history bit-string. In the next
occurrence of the same history bit-string, the virtual points
awarded now lead to a crowd behavior \cite{johnson3} with a crowd
of agents deciding on the same outcome as in the last occurrence.
However, for small $m$, this crowd tends to be too big to win, due
to the small strategy space and hence substantial overlap of
strategies among agents. Hence, the outcome is opposite to that in
the last occurrence of the same history bit-string. Virtual points
are then awarded.  For the next occurrence of the same bit-string,
the situation is similar to that of the first occurrence.  Since
there are a total of $2^m$ possible history bit-strings, it takes
on average $2\times 2^m$ time steps to sample all the history
bit-strings twice, thus leading to the doubly-periodic features.
In graph-theoretical language, this is related to the path in the
de Bruijn graph formed by the possible history bit-strings. A
crowd-anticrowd theory \cite{johnson3,hart1} can be formulated to
explain the large standard deviation $\sigma$ in the number of
agents making a particular decision over time for small $m$ in
terms of the crowd effect.  A large fluctuation implies fewer
winners per turn and hence a low success rate. The success rates
shown in Fig.1 for the other agents is strongly correlated with
$\sigma$. For the RPA, the virtual merit points of her strategies
are different from that of the other agents, as she rewards the
strategies only in the turns she participates.  Therefore, the RPA
does not fully adapt to the information created by the other
agents, and hence does not become part of the crowd.  This gives
the RPA the ability to avoid over-adaptation in the crowded phase
and an winning edge over the other agents. We have checked that if
the random agent keeps record of the performance of her strategies
for the turns that she does not enter, the strategies still adapt
to the global history and no enhanced success rate results, even
she participates randomly.

\section{Inefficiency and success rate}
\subsection{Numerical Results}

To explore deeper into the underlying physics for the enhanced
success rate of the RPA, we study the statistics of the
bit-strings in the series consisting of the outcomes for the turns
that the RPA has participated. This series of bit-strings
corresponds to the one with which the strategies of the RPA are
assessed for their performance. To quantify our discussion, we
focus on the behavior in the range of small $m$ and look at the
probability of a winning outcome of `1' following a given
bit-string of $m$-bits. We define the inefficiency $\varepsilon$
\cite{challet3,challet5,challet6} as follows:
\begin{equation}
\varepsilon = \frac{1}{2^m} \sum_{i=1}^{2^{m}}
|P(1|i(m)) - 1/2|,
\end{equation}
where the sum is over all $2^{m}$ possible $m$-bit strings and
$P(1|i(m))$ is the conditional probability that a given $m$-bit
string labelled by $i$ is followed by an outcome `1' in a long
series of the RPA bit-strings in a particular realization of the
model \cite{notes1}. The inefficiency $\varepsilon$ measures the
information left in the history bit-strings that the RPA uses to
assess her strategies. Note that for the MG in the crowded phase,
$\varepsilon = 0$ \cite{savit}, indicating that there is no
information left in the history bit-strings.

Each realization (run) of the model gives one value of
$\varepsilon$.  Figure 3 shows the distribution of the values of
$\varepsilon$ over 10,000 independent runs with different initial
conditions for $m=2$ and $m=4$ (inset) and $q=0.5$. Strikingly,
$\varepsilon \neq 0$ in general for the bit-strings specific to
the RPA. The distribution shows a peak at finite $\varepsilon$
with the most probable value of $\varepsilon$ increases with $m$.
We have checked that the inefficiency of the bit-string in a run
used by the other agents takes on a value very close to zero.

From Fig.1 and Fig.3, there is a spread of success rate for
different runs that is associated with a spread in the values of
$\varepsilon$.  For each run of our model, there is a success rate
$R$ and a value of $\varepsilon$. It is therefore interesting to
explore the correlation, if any, between $R$ and $\varepsilon$ in
a run. Figure 4 shows a plot of the success rate $R$ against the
inefficiency $\varepsilon$ for a system with $N=101$, $s=2$, $m=2$
and a participating probability $q = 0.5$ for the RPA.  Each data
point on the plot represents the success rate and the value of
inefficiency in a run. There are 10,000 data points on the plot,
corresponding to 10,000 independent runs.  From the scattered data
points, there {\em emerges} a straight line consisting of a
fraction of $\sim 0.06$ of the total number of runs \cite{notes2}.
For these runs, the success rate $R$ is related to the
inefficiency $\varepsilon$ by $R = 0.5 - \varepsilon$. The inset
of Fig.4 shows the results for $m=4$ in which the RPA holds two
identical strategies, the implication of which will be discussed
in the next subsection.

\subsection{Theoretical Analysis}

The emergence of the relationship between the success rate and the
inefficiency can be understood by invoking the doubly-periodic
feature in MG, as described in the last section. Let
$t_{even}^{\mu }$ ($t_{odd}^{\mu }$) be a set consisting of the
turns  in a history series that a particular history $\mu $
occurred an even (odd) number of times from the beginning of the
game just before the moment of decision with history $\mu$. Let
$P(\tau |t_{even}^{\mu })$ ($P(\tau |t_{odd}^{\mu })$) be the
probability that the outcome is $\tau$ ($\tau =0,1$) at a turn $t$
that belongs to $t_{even}^{\mu }$ ($t_{odd}^{\mu }$).
Double-periodicity implies $P(\tau |t_{even}^{\mu })= \frac{1}{2}$
and $P(\tau |t_{odd}^{\mu })=1-\tau.$ To proceed, we note that the
straight line in main panel of Fig.4 consists of about 600 data
points out of a total of 10,000 runs.  This amounts to a fraction
of $1/(2^{2^{m}}) = 1/16$ for $m=2$, which is the fraction of
agents holding two identical strategies in a population.
Therefore, the runs on the straight line corresponds to the cases
in which the RPA picks two identical strategies, i.e., effectively
one strategy.  This is further confirmed numerically in the inset
of Fig.4 in a game of $m=4$ in which the RPA is restricted to have
two identical strategies.  We further note that it is the RPA that
generates the inefficiency $\varepsilon$.  This is related to the
fact that, unlike an outsider who observes but not participates,
an agent's action affects the outcome, and hence her own success
rate.  This is the agent's ``market impact".  Consider, for
example, the case that the other ($N-1$) agents are equally split
between the two options.  The outcome will always be the opposite
of that of the remaining agent. For a RPA, however, this market
impact effect is important only for the turns $t\in t_{even}^{\mu
}$.  For $t\in t_{odd}^{\mu }$, the difference in the number of
agents taking the two options is so large for a single RPA to
affect the outcome.

Consider a RPA with two identical strategies.
If the agent decides to enter the game in a turn $t\in t_{even}^{\mu }$
there is a small probability $P$ in these turns that the other
agents are evenly divided between option ''$0"$ and option ''$1"$.
In this case, the outcome is determined by the strategy of the
RPA. If her strategy predicts option $\tau $, the outcome will be
$(1-\tau )$ and the agent will definitely lose. In the next
occurrence of the history $\mu$, the outcome must be $\tau $ due
to the doubly-periodic feature. As a result, the RPA will
definitely win if she decides to participate. The success rate of
the agent is then given by \cite{larry}
\begin{eqnarray}
R &=&(1-P-qP)\cdot 1/2+P\cdot 0+qP\cdot 1  \nonumber \\
&=&\frac{1}{2}-\frac{1-q}{2}P ,
\end{eqnarray}
where the last two terms in the first line correspond to the two
cases discussed above, and the factor $(1-P-qP)$ is the
probability of occurrence of cases other than those included in
the last two terms.

The inefficiency $\varepsilon $ can be calculated using Eq.(1) in
a similar manner. There is a probability $P$ in the turns that the
RPA participates that the RPA must lose for $t\in t_{even}^{\mu
}$, with an outcome $(1-\tau)$. If the agent participates in the
next occurrence of that particular history $\mu$, the outcome will
be $\tau$. Consequently, the probability for the RPA to observe an
outcome $\tau $ is
\begin{eqnarray}
P_{\tau } &=&(1-P-qP)\cdot 1/2+P\cdot 0+qP\cdot 1  \nonumber \\
&=&\frac{1}{2}-\frac{1-q}{2}P ,
\end{eqnarray}
where $\tau$ takes on either ``$0$" or ``$1$".
It follows from Eq.(1) that
\begin{eqnarray}
\varepsilon  &=&\frac{1}{2^{m}}\sum_{\mu =1}^{2^{m}}|P(1|t^{\mu })-1/2|  \nonumber \label{num4}
\\
&=&\frac{1-q}{2}P .
\end{eqnarray}%
Combining Eqs.(2) and (4), we have
\begin{equation}
R=\frac{1}{2}-\varepsilon ,
\end{equation}
as observed numerically for runs in which the RPA holds two
identical strategies.

\section{One Randomly participating Agent: $\lowercase{q}\neq 0.5$}

Figure 5 shows the results of the success rate, averaged over 50
runs for each data point, of a RPA as a function of $m$ for a
system with $N=101$, $s=2$ for different values of participation
probability $q$ ($q=0.3,0.5,0.8$).  For $q=1$, the success rate of
the RPA is the same as that of the other agents and the results
are identical to that of the MG.  It is noted that for small
values of $m$, i.e., in the crowded phase, the enhanced success
rates for $q\neq 1$ take on similar values, all of which are
significantly higher than that of the other agents (reasonably
represented by the $q=1$ results).  For values of $m$ in the
inefficient phase of the MG, a higher participation probability
gives a higher success rate.  However the success rate is still
lower than that of the other agents.

The argument in the previous section can be readily generalized to
arbitrary values of $q \neq 1$.  In this case, the success rate of
a RPA holding repeated strategies is given by
\begin{equation}
R(q) = \frac{1}{2} - \varepsilon(q),
\end{equation}
i.e, taking on the same form as Eq.(5). The inefficiency
$\varepsilon(q)$, similar to Eq.(4), is given by
\begin{equation}
\varepsilon(q) = \frac{1-q}{2} P(q),
\end{equation}
where $P(q) = T_{c}(q)/T$ with $T_{c}(q)$ being the number of
turns in a run of $T$ turns ($T \gg 1$) that the other $(N-1)$
agents are evenly split between the two decisions {\em and} the
RPA participates.  The result suggests that for different values
of $q$, a plot of the success rate $R$ against $\varepsilon(q)$
will have data points clustered on a straight line with slope
-$1$. This is indeed the case as shown for data obtained with 5
different values of $q$ in Fig.6.  For each value of $q$, 500
independent runs are carried out, and there are 2,500 data points
on the plot. The data clearly exhibit the behavior suggested by
Eq.(6).

It is interesting to explore the distribution of inefficiency for
values of $q$ away from $q=1$.  Figure 7 shows the probability
density of inefficiency for four different values of $q$
($q=0.3,0.5,0.7,0.9$) in a system with $N=101$, $m=4$ and $s=2$.
For each value of $q$, the distribution is obtained from 1,000
runs. It should be pointed out that the distribution for the basic
MG ($q=1$) is sharply peaked near $\varepsilon=0$. As $q$ takes on
values gradually away from unity, the distribution of inefficiency
gradually spreads wider with the most probable value of the
inefficiency increases as the deviation of $q$ from unity
increases. Therefore, the enhanced success rate is accompanied by
an non-vanishing inefficiency in the range of small $m$.

\section{Many Randomly Participating Agents}

It is important to investigate whether the enhanced success rate
of a RPA persists when more RPAs are present in a population.
Figure 8 shows the averaged success rate as a function of the
fraction, $N_{RPA}/N$, of RPAs in a population with $N=101$,
$s=2$, $m=2$ and $q=0.5$.  The dashed line gives the success rate
of the RPAs and the solid line gives the success rate of the other
agents, for given $N_{RPA}/N$. Most strikingly is that for a wide
range of $N_{RPA}/N$ up to about 60\%, the success rates of the
RPAs and the rest of the population are basically insensitive to
$N_{RPA}$. Over this range of $N_{RPA}/N$, the success rate of the
RPAs are much higher than that of the other agents, as in the case
of $N_{RPA}=1$.  The results show that the system is still
dominated by the crowd effect for $N_{RPA}/N < 0.6$, and the RPAs
achieve a higher success rate by avoiding themselves from the
crowd. For $N_{RPA}/N \geq 0.6$, the success rate of the RPAs
decreases with $N_{RPA}/N$ and that of the other agents increases.
It is expected that the fraction of RPA over which the RPAs
out-perform the other agents may depend on the value of $q$.  For
$q=0.5$ discussed here, half of the RPAs participate in each turn
on the average.  Thus the number of participating agents
effectively reduces.  Up to a fraction of 60 \%, the majority of
the participating agents in a turn are the other agents.  Thus,
the system is still influenced by the actions of the other agents.

\section{Conclusion}

We considered the success rate of a number of randomly
participating agents in a population competing for limited
resource. The RPAs participate with a probability $q$ and they
assess the performance of their strategies only in the turns that
they participate.  Extensive numerical calculations were carried
out.  For one RPA in a population, it was found the RPA has a
higher success rate than the other agents in the crowded phase.
Rewarding the strategies only in the turns of participation avoids
the RPA from over-adaptation to the outcomes produced by the
actions of the other agents.  The hidden information in the
bit-strings that a RPA used to reward her strategies was analyzed
in terms of the inefficiency $\varepsilon$. In basic MG,
$\varepsilon=0$ for all agents. In our model, the RPA sees
$\varepsilon>0$.  Numerical data reveals a relationship between
the success rate $R$ and $\varepsilon$ of the form $R = 0.5
-\varepsilon$, if the RPA holds two identical strategies. The
relationship was explained in terms of the dynamics through which
the possible history bit-strings were visited. Results for $q \neq
0.5$ were also presented and discussed.  The RPA has an enhanced
success rate for $q \neq 1$, together with non-vanishing
$\varepsilon$. For many RPAs in a population, it was found that
the winning edge of the RPAs persists even for a population up to
about 60\% of RPAs for $q=0.5$.

It is also interesting to explore if it is random participation
alone that leads to the enhanced success rate.  We, therefore,
consider the case of a fraction of agents $N_{random}/N$ who
participate with probability $q$ {\em and} decide the action
randomly, i.e., they do not use the most recent $m$ outcomes for
decisions. Results are shown in Fig.8 for comparison. The success
rate of these random agents (dot-dashed line), together with that
of the other agents (solid line with symbols) are shown as a
function of $N_{random}/N$.  The results show that the success
rate of the RPAs is {\em consistently higher} than that of the
random agents for all ratio in the population. We have also
checked that the RPAs perform better than random agents for all
values of $m$.  Thus, while random participating does lead to a
better performance by avoiding the crowd, the scheme of rewarding
virtual points to the RPAs and making use of the virtual points in
decision making gives an additional advantage for a further
improved performance.

\begin{acknowledgements}
This work was supported in part by the Research Grants Council of
the Hong Kong SAR Government under grant number CUHK 4241/01P.
\end{acknowledgements}

\newpage \centerline{\bf Figure Captions}

\bigskip

\noindent Figure 1: The success rates $R$ of the randomly
participating agent and the other ($N-1$) agents as a function
of $m$ for $N=101$, $q=0.5$ and for (a) $s=2$, and (b) $s=3$.
For each value of $m$,
the data correspond to
50 independent realizations.
The lines represent an average over 50 runs. For $m < m_{o}$, an
the RPA performs significantly better than
the other agents.

\bigskip

\noindent Figure 2: The success rate $R$ of the
randomly participating agent as a function of $m$
for three different systems: $q=0.5$, $N=101$ with $s=2$ and
for $N=301$ with $s=2,3$. Each data point represents
an average over 50 independent realizations.

\bigskip

\noindent Figure 3: The probability density of inefficiency
$\varepsilon$ for $N=101$, $m=2$, $s=2$ and $q=0.5$. The inset
shows the probability distribution of $\varepsilon$ for $m=4$.
Each distribution is obtained from the results of 10,000
independent runs.

\bigskip

\noindent Figure 4: The success rate $R$ against $\varepsilon$
in the bit-string with which the RPA uses to assess
the performance of her strategy.  The parameters are $N=101$, $s=2$, $m=2$
and $q = 0.5$ for the RPA.
There are 10,000 data points on the plot corresponding to 10,000
independent runs.  Note that the data points lead to the emergence
of a line.  The inset shows
the results for $m=4$ in which the RPA holds two identical strategies.

\bigskip

\noindent Figure 5: The average success rate of the RPA
as a function of $m$ for a system with $N=101$,
$s=2$ for different values of $q=0.3, 0.5, 0.8, 1$.
Each data point is an average over 50 independent runs. The lines
are guide to eye.

\bigskip

\noindent Figure 6: The success rate $R$ against $\varepsilon$
with $N=101$, $s=2$ and $m=2$ for five different
values of $q=0.3, 0.4, 0.5, 0.6, 0.7$.  The RPA holds
two identical strategies.  For each value of $q$ (labelled by a symbol),
500 independent runs are carried out.  The solid line represents
the relationship $R = 0.5 -\varepsilon$.

\bigskip

\noindent Figure 7: The probability density of inefficiency
$\varepsilon$ for the RPA in a system with $N=101$, $s=2$, $m=2$
for four different values of $q=0.3, 0.5, 0.7, 0.9$. Other
parameters are $N=101$, $m=4$ and $s=2$.  For each value of $q$,
the distribution is obtained from 1,000 independent runs.

\bigskip

\noindent Figure 8: The average success rate as a function of the
ratio $N_{RPA}/N$ of RPAs in the population with $N=101$, $s=2$, $m=2$
and $q=0.5$.  A pair of lines show the success rate of the RPAs (dashed line)
and the other agents (solid lines).  For comparison, another pair of lines
show the success rate of random agents (dot-dashed line) and the
other agents (solid line with symbols) as a function of the ratio
$N_{random}/N$.  Each data point represents an average over 50
independent realizations.

\newpage
\begin{figure}
\epsfig{figure=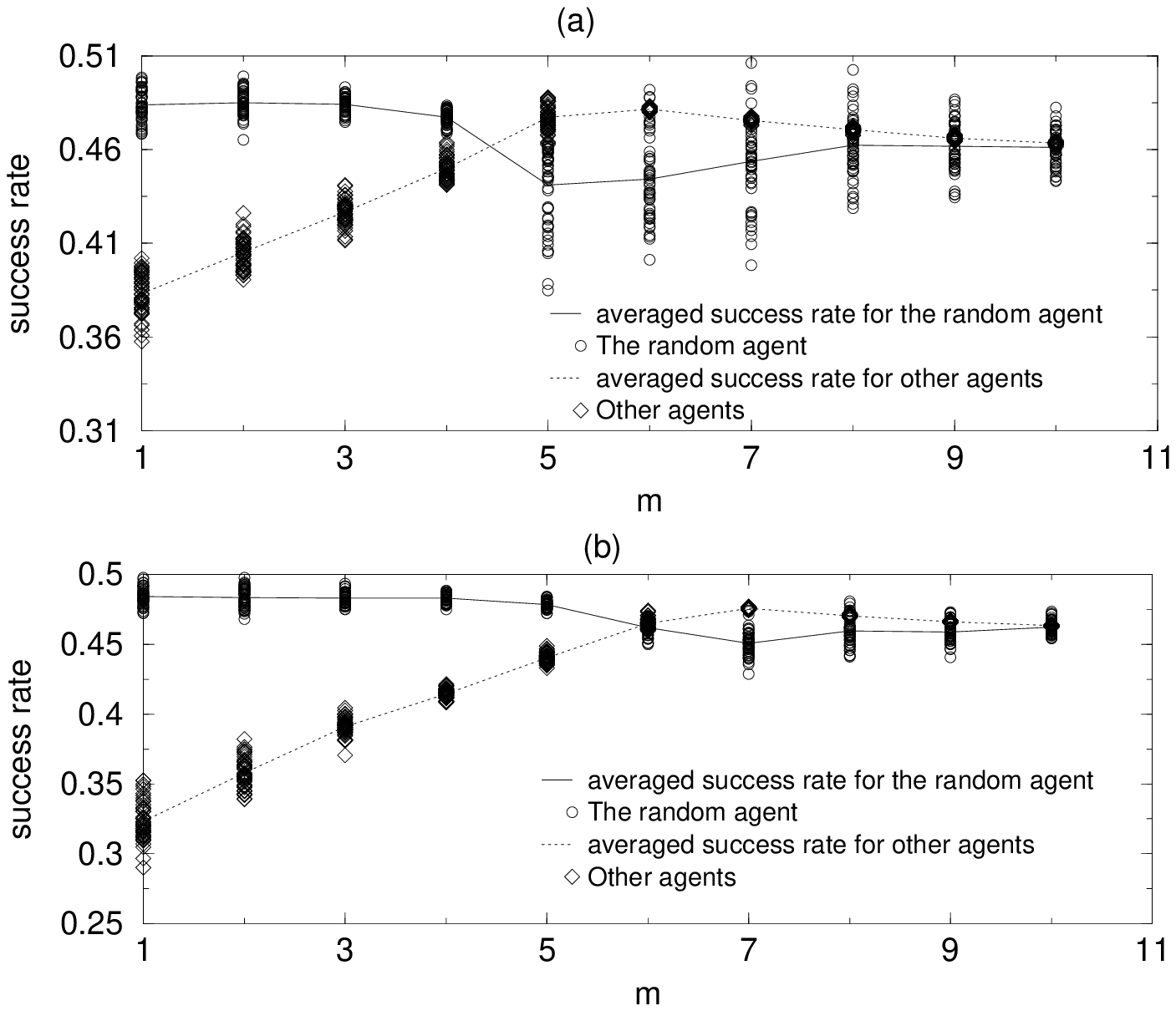,width=\linewidth}
\label{figure1}
\end{figure}

\newpage
\begin{figure}
\epsfig{figure=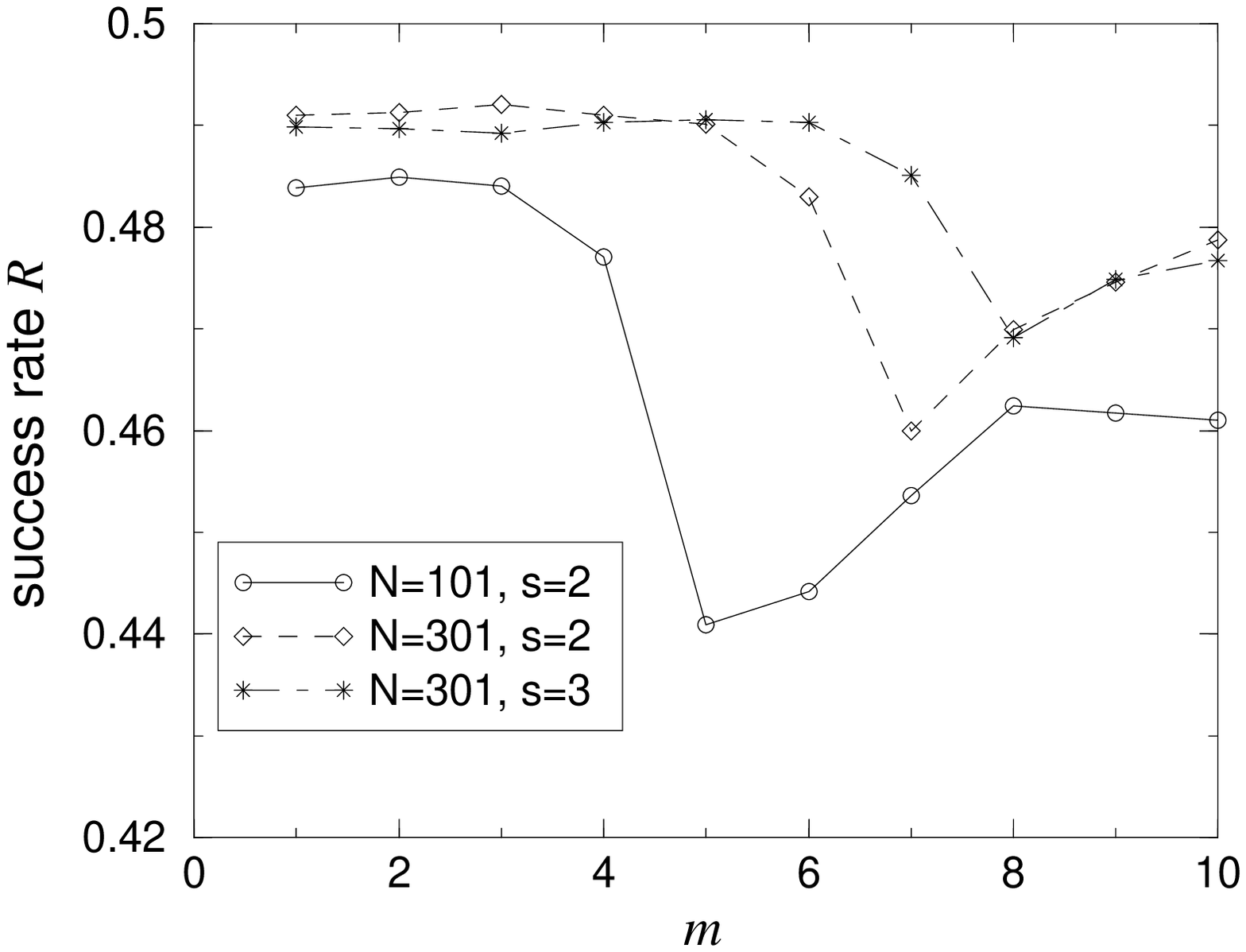,width=\linewidth}
\label{figure2}
\end{figure}

\newpage
\begin{figure}
\epsfig{figure=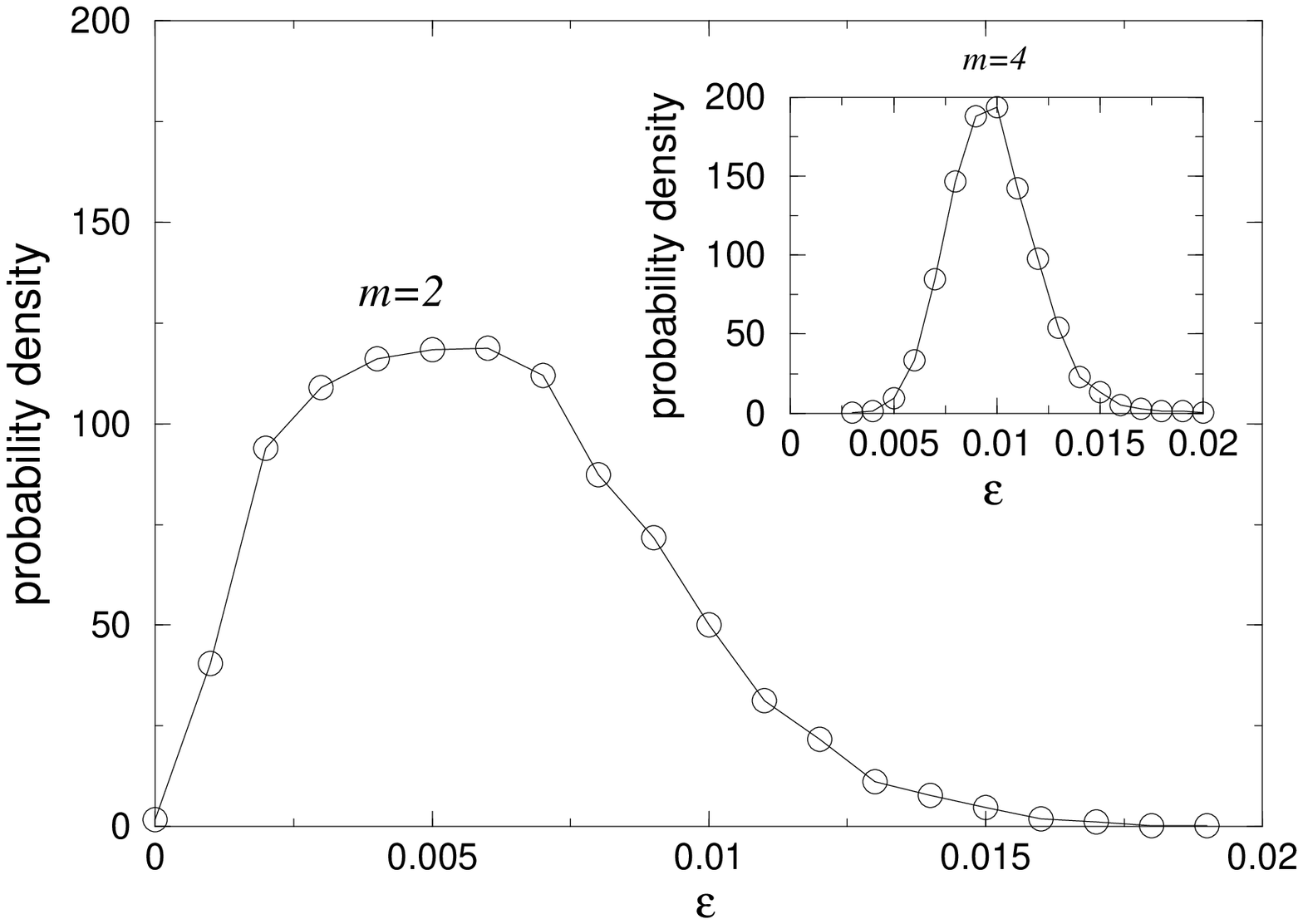,width=\linewidth}
\label{figure3}
\end{figure}

\newpage
\begin{figure}
\epsfig{figure=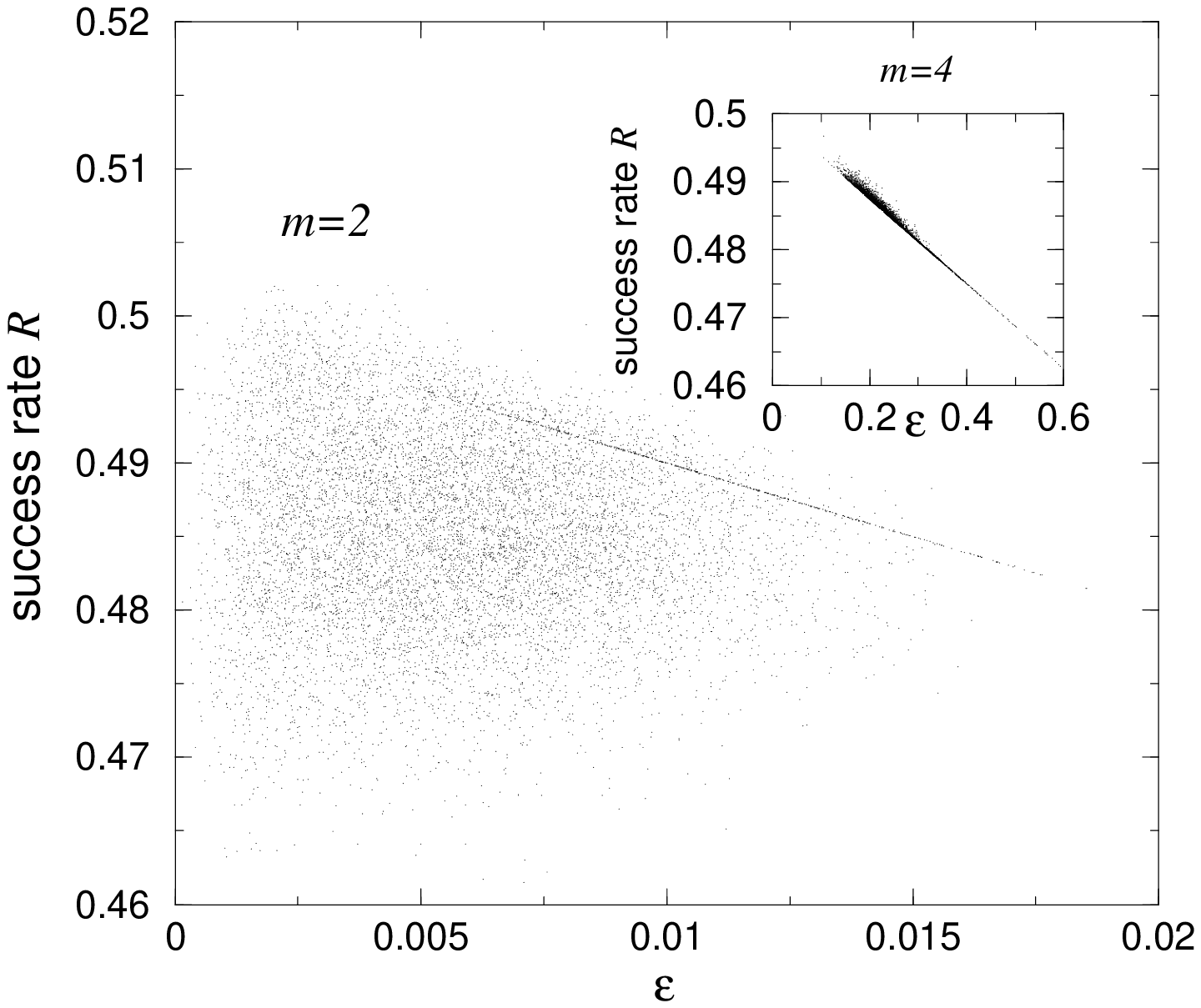,width=\linewidth}
\label{figure4}
\end{figure}

\newpage
\begin{figure}
\epsfig{figure=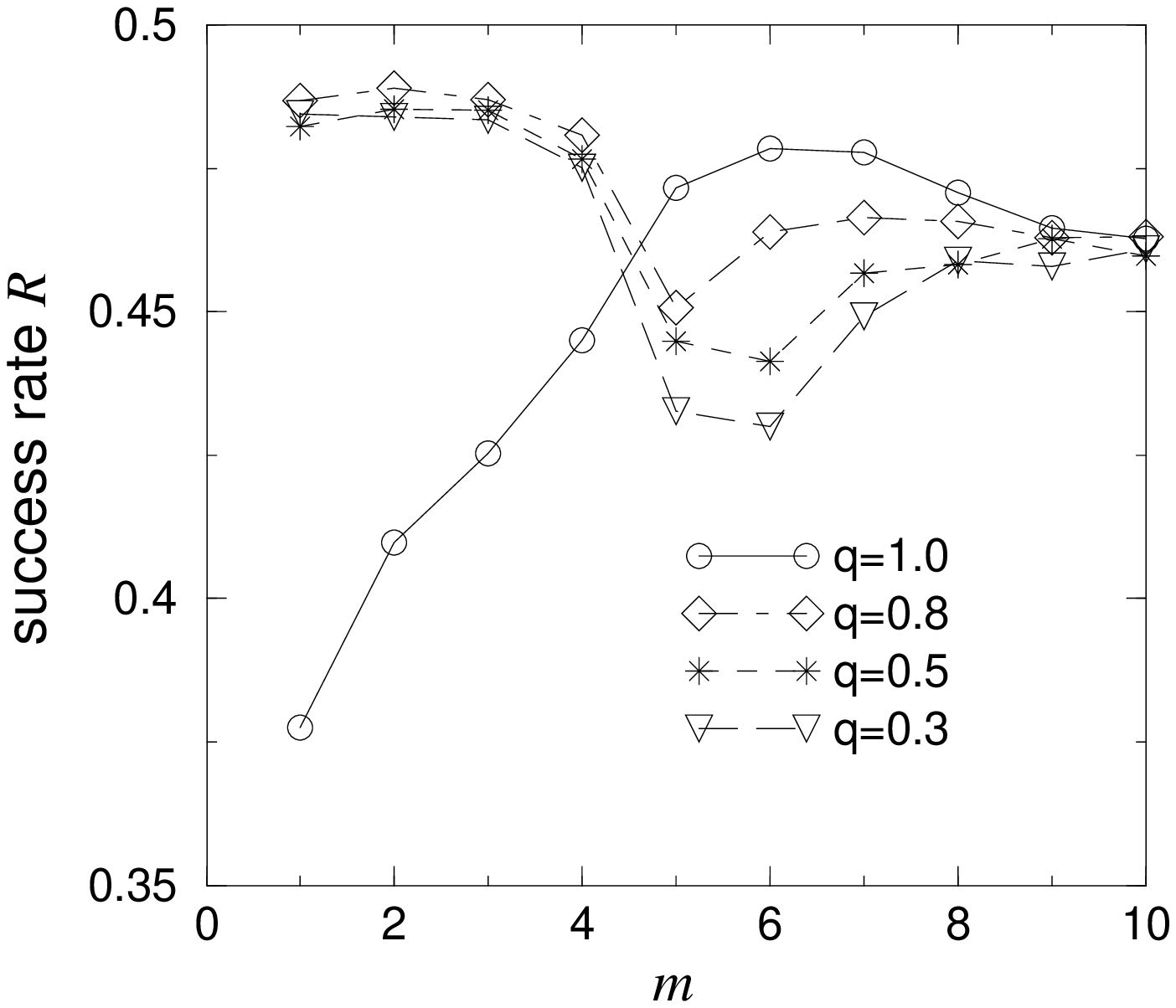,width=\linewidth}
\label{figure5}
\end{figure}

\newpage
\begin{figure}
\epsfig{figure=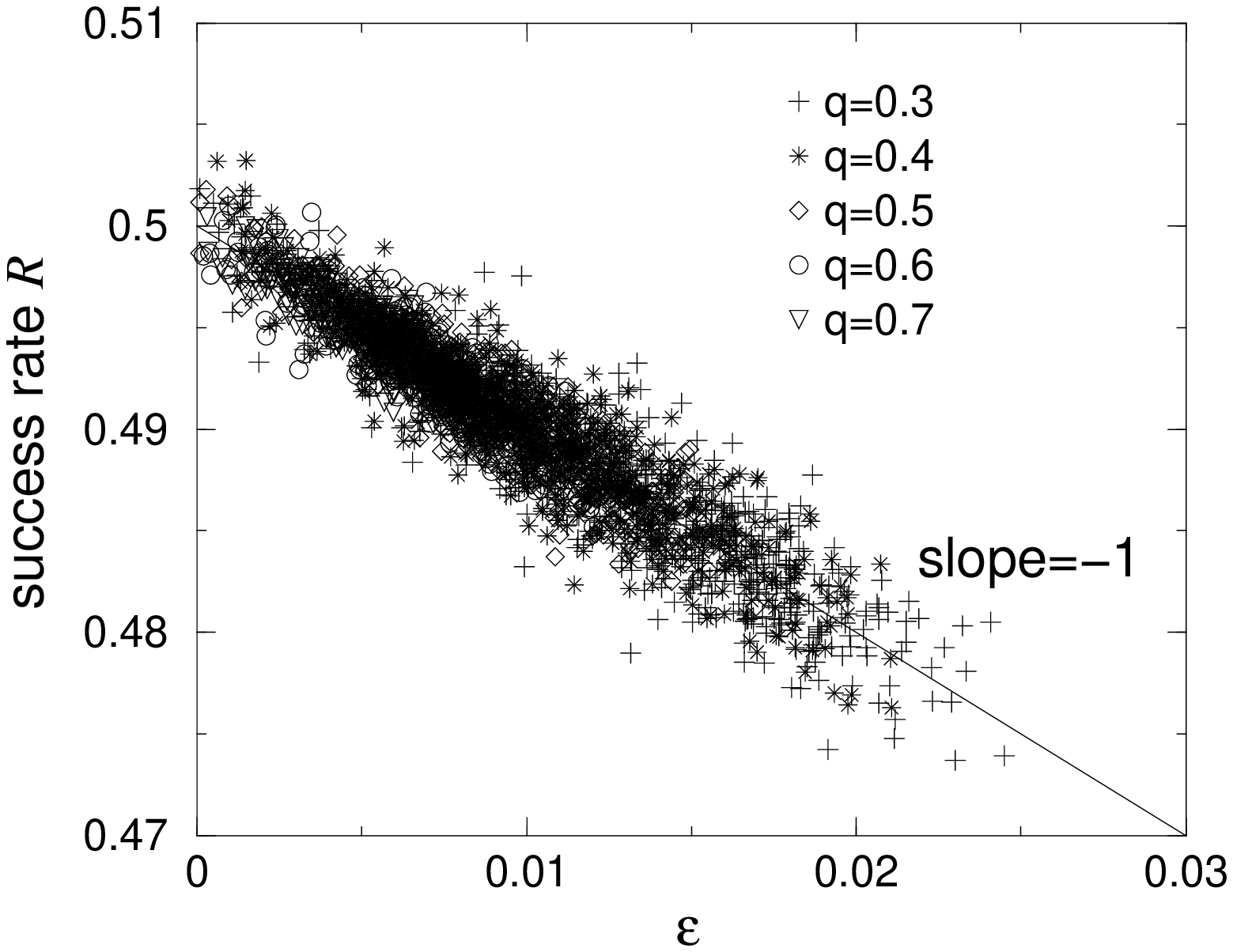,width=\linewidth}
\label{figure6}
\end{figure}

\newpage
\begin{figure}
\epsfig{figure=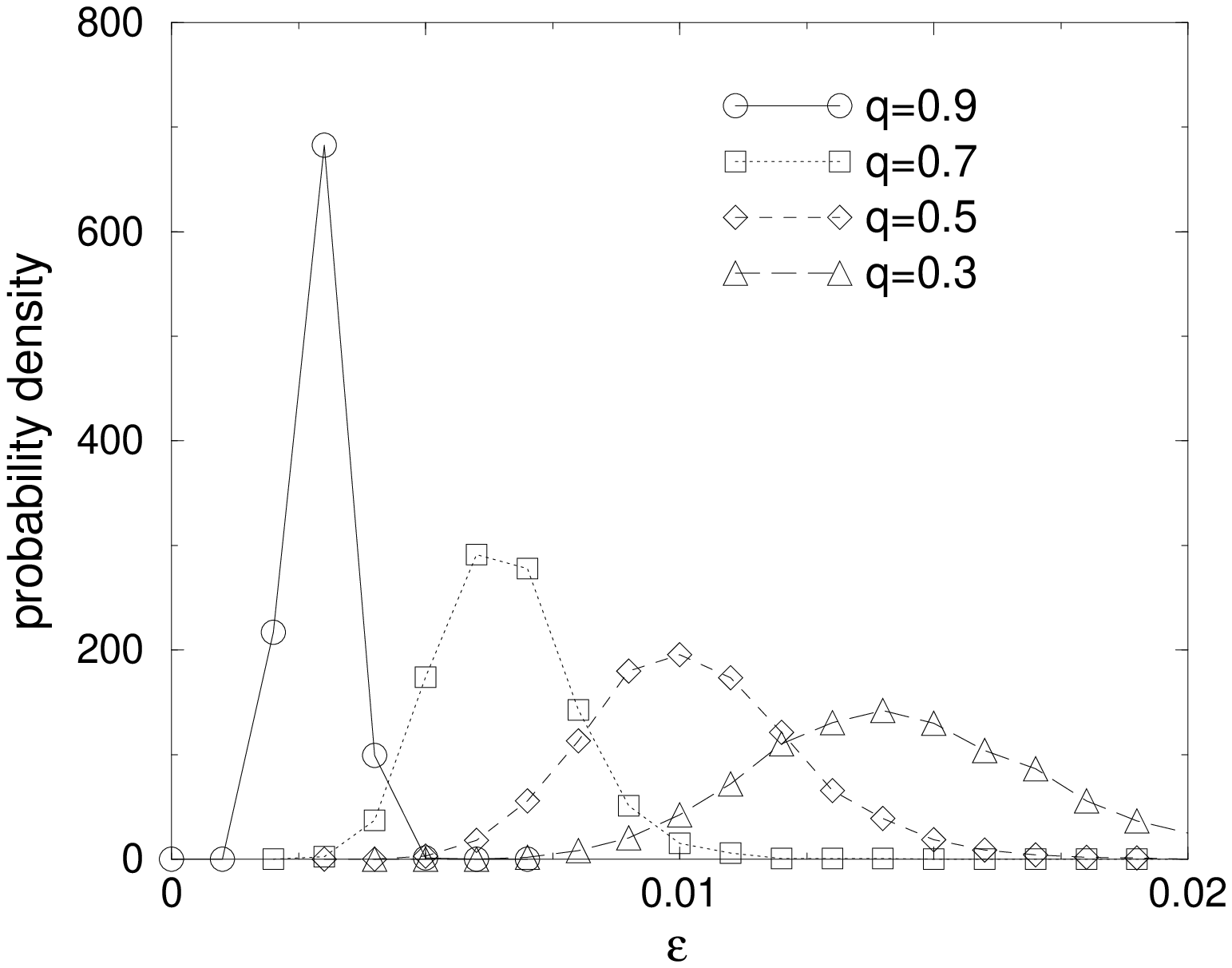,width=\linewidth}
\label{figure7}
\end{figure}

\newpage
\begin{figure}
\epsfig{figure=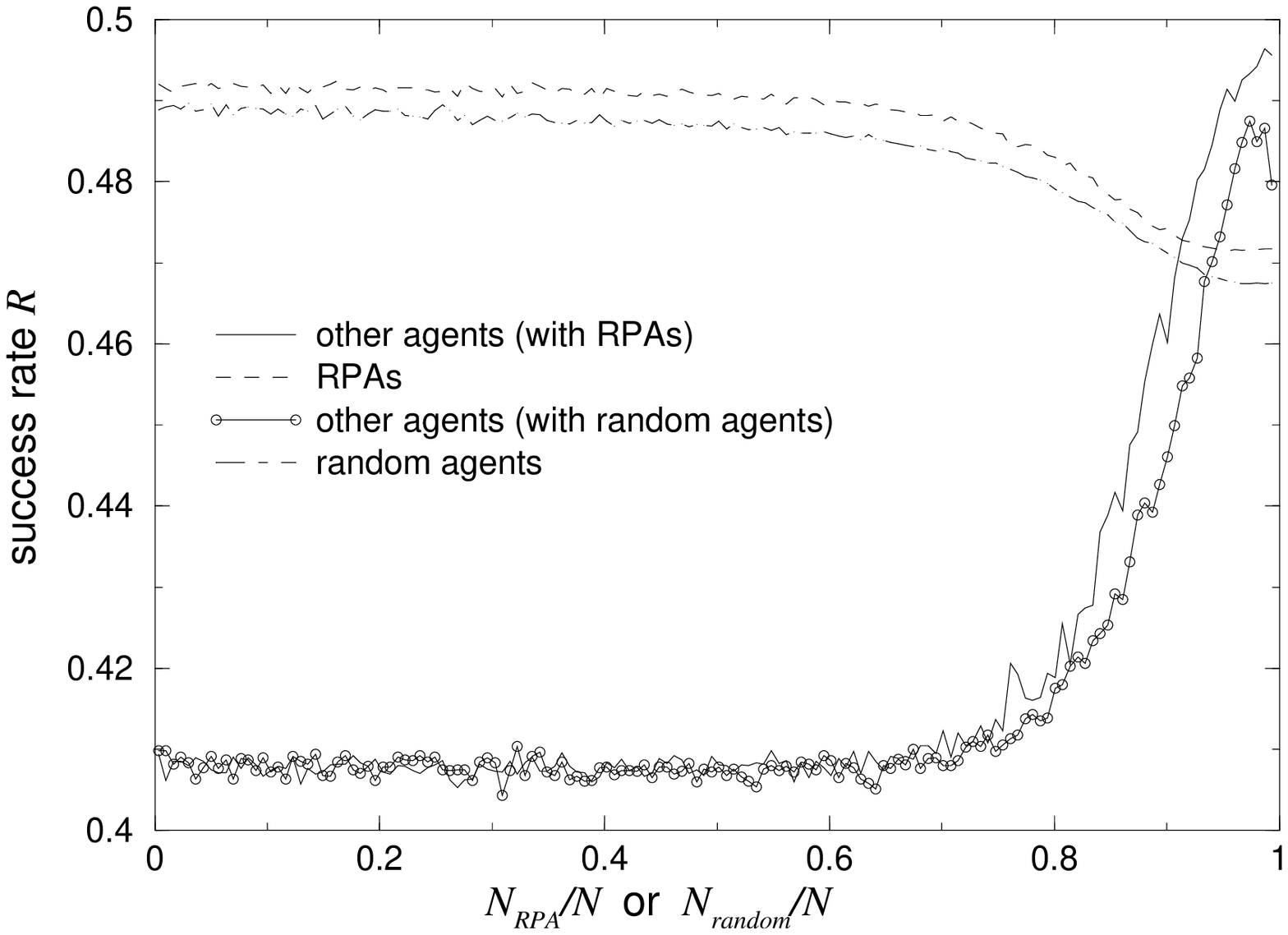,width=\linewidth}
\label{figure8}
\end{figure}


\begin{thebibliography}{99}
\bibitem{notes1} The definition of $\varepsilon$ in Eq.(1)
is similar to the variables $\theta$ defined in Ref.\cite{challet5}
and $H$ in Ref.\cite{challet6} in analyses of the MG
within a spin-glass-type formalism.
\bibitem{notes2} The full strategy space with $2^{2^{m}}$ strategies
is used in the numerical studies.  If a reduced strategy space is used,
the fraction of points on the emerged straight line is about 12.6\%.
\end{thebibliography}
\end{document}